\documentclass[%
reprint,
superscriptaddress,
amsmath,amssymb,
prb,
longbibliography
]{revtex4-2}

\usepackage{color}
\usepackage{xcolor}
\usepackage{graphicx}
\usepackage{dcolumn}
\usepackage{siunitx}
\usepackage{bm}
\usepackage{hyperref}
\hypersetup{colorlinks=true,linkcolor=blue,citecolor=blue,urlcolor=blue}
\usepackage{soul} 
\usepackage[colorinlistoftodos]{todonotes}
\usepackage{mhchem}
\usepackage{textcomp}

\begin{document}

\title{Hybridization-Induced Enhancement of the Thermoelectric Response in the Kagome Material CsCr$_6$Sb$_6$}

\author{Hongye Zhang}
\thanks{These authors contributed equally to this work.}
\affiliation{Low Temperature Physics Laboratory, School of Physics \& Center of Quantum Materials and Devices, Chongqing University, Chongqing 401331, China}

\author{Boqin Song}
\thanks{These authors contributed equally to this work.}
\affiliation{Beijing National Laboratory for Condensed Matter Physics, Institute of Physics, Chinese Academy of Sciences, Beijing 100190, China}

\author{Aifeng Wang}
\affiliation{Low Temperature Physics Laboratory, School of Physics \& Center of Quantum Materials and Devices, Chongqing University, Chongqing 401331, China}

\author{Yisheng Chai}
\affiliation{Low Temperature Physics Laboratory, School of Physics \& Center of Quantum Materials and Devices, Chongqing University, Chongqing 401331, China}

\author{Xiaoyuan Zhou}
\affiliation{Low Temperature Physics Laboratory, School of Physics \& Center of Quantum Materials and Devices, Chongqing University, Chongqing 401331, China}

\author{Ying Zhu}
\email{20242701033@stu.cqu.edu.cn}
\affiliation{Low Temperature Physics Laboratory, School of Physics \& Center of Quantum Materials and Devices, Chongqing University, Chongqing 401331, China}

\author{Tianping Ying}
\email{ying@iphy.ac.cn}
\affiliation{Beijing National Laboratory for Condensed Matter Physics, Institute of Physics, Chinese Academy of Sciences, Beijing 100190, China}

\author{Mingquan He}
\email{mingquan.he@cqu.edu.cn}
\affiliation{Low Temperature Physics Laboratory, School of Physics \& Center of Quantum Materials and Devices, Chongqing University, Chongqing 401331, China}

\date{\today}

\begin{abstract}
Flat bands in kagome lattices provide a prominent route to strongly correlated electronic states. Here we report charge and thermal transport measurements on single crystals of \ce{CsCr6Sb6}, a kagome Kondo system arising from the hybridization between flat and dispersive bands. The Seebeck coefficient and Nernst signal increase rapidly below $T_\mathrm{F}\sim70$ K, concomitant with the emergence of short-range antiferromagnetic correlations and flat-band resonance. The Nernst coefficient in the intermediate-temperature range of $\sim30$--80 K further follows the universal scaling reported for prototypical heavy-fermion compounds, consistent with the development of strongly renormalized quasiparticles associated with hybridization. Both the Seebeck and Nernst responses reach extrema at a lower temperature $T_\mathrm{p}\sim20$--25 K, where phonon drag and correlated quasiparticles may both contribute. These results reveal a hybridization-induced enhancement of the thermoelectric response and establish \ce{CsCr6Sb6} as a kagome platform for probing flat-band-derived Kondo physics through thermoelectric transport.
\end{abstract}

\maketitle

The kagome lattice provides a paradigmatic platform for studying intertwined lattice geometry and electronic band topology \cite{Syozi1951}. Its electronic structure naturally hosts Dirac cones, van Hove singularities, and flat bands, enabling a wide range of correlated and topological phases \cite{Brien2010,ko2009doped,Balents2010,kiesel2012sublattice,Kisel2013,Yu2012,Ghimire2020,Neupert2022,Jiang2023}. When a flat band approaches the Fermi level, the enhanced density of states (DOS) may promote Anderson- or Kondo-lattice physics, exhibiting unconventional quantum critical behavior \cite{Chen_kondo2024,Chen_kondo,Wang_kondo}. While kagome materials have already yielded quantum spin-liquid candidates \cite{Helton_QSL,Han2012_QSL}, topological metals \cite{Nakatsu_topo,Ye2018_topo,Liu2018_topo,Kang2020_topo}, and superconductors \cite{AV3Sb5Prototype,CsV3Sb5SC,KV3Sb5SC,KV3Sb5CDW}, experimental realizations of Kondo physics in kagome systems remain rare. Recent observations of flat-band-induced Kondo signatures in \ce{Ni3In} and pressurized \ce{CsCr3Sb5} have begun to reveal this largely unexplored regime \cite{Ye2024_Nb3In,Liu2024_CsCr3Sb5}.

The kagome compound \ce{CsCr6Sb6} offers a particularly clean opportunity to access such physics at ambient pressure. In contrast to many kagome metals where flat bands lie away from the Fermi energy $E_\mathrm{F}$, bulk \ce{CsCr6Sb6} hosts flat bands in close proximity to $E_\mathrm{F}$ \cite{CsCrSbNC}. Their hybridization with dispersive itinerant bands opens a gap of approximately 10--15 meV, giving rise to a crossover from metallic behavior at high temperatures to Kondo-insulating behavior at low temperatures \cite{CsCrSbNC}. Subsequent spectroscopic measurements further reveal a hybridization-induced flat-band resonance and Kondo resonance features \cite{CsCrSbPRB,CsCrSbLiu2025,Zhang2026_CsCr6Sb6}. Notably, the incoherence-to-coherence crossover of the  flat-band resonance occurs around $T_\mathrm{F}\sim70$ K, concomitant with the emergence of short-range antiferromagnetic correlations \cite{CsCrSbLiu2025,Zhang2026_CsCr6Sb6}. Thus, hybridization, magnetic correlations, and Kondo screening are closely intertwined in this flat-band-derived kagome system \cite{CsCrSbNC,CsCrSbPRB,CsCrSbLiu2025,Zhang2026_CsCr6Sb6}.  

Thermoelectric transport is exceptionally sensitive to this kind of hybridization-driven low-energy reconstruction. Unlike the longitudinal electrical resistivity, which mainly tracks momentum relaxation, the Seebeck and Nernst responses probe carrier entropy, mobility imbalance, and the energy dependence of the conductivity near $E_\mathrm{F}$ \cite{behnia2009,behnia2016}. In Kondo and narrow-gap correlated systems, these quantities can be strongly amplified around the crossover from incoherent local-moment scattering to coherent  quasiparticle transport.


Here we use thermoelectric transport to probe the incoherence-to-coherence crossover associated with hybridization in single-crystalline \ce{CsCr6Sb6}. Both the Seebeck coefficient and Nernst signal begin to increase rapidly below $T_\mathrm{F}$, where the coherent flat-band resonance and short-range magnetic correlations emerge. In particular, the Nernst coefficient in the intermediate-temperature regime follows the universal scaling reported for various heavy-fermion systems, consistent with a hybridization-induced asymmetry of the quasiparticle DOS. Both the Seebeck and Nernst coefficients develop broad extrema near $T_\mathrm{p}\sim20$--25 K and respond only weakly to magnetic fields up to 9 T, likely suggesting sizable phonon-drag contributions.

\begin{figure*}
\centering  
\includegraphics[scale=0.35]{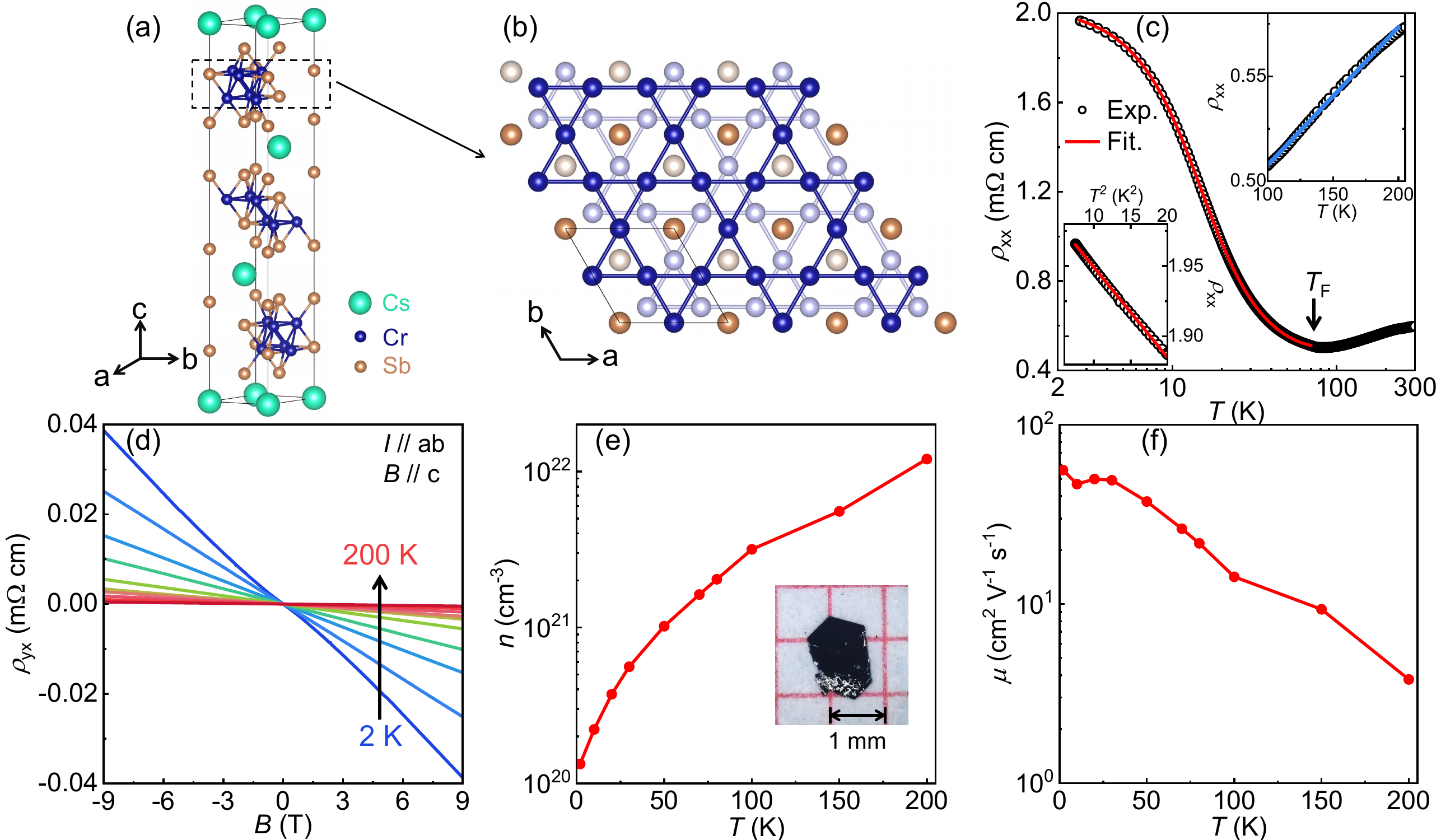}
\caption{(a) Side view of the crystal structure of \ce{CsCr6Sb6}. (b) Top view of the Cr double-kagome layer along the $c$ axis. (c) Temperature dependence of the longitudinal electrical resistivity $\rho_{xx}$ measured at zero magnetic field. The red curve is a fit to $\rho_{xx}=\rho_0+AT+\rho_\mathrm{K}$, where the Kondo contribution $\rho_\mathrm{K}$ is described by the modified Hamann expression in Eq.~(\ref{eq:rhoK}). The upper-right inset shows the nearly linear temperature dependence of $\rho_{xx}$ between 100 and 200 K. The lower-left inset highlights the $-T^2$ dependence of $\rho_{xx}$ at low temperatures. (d) Hall resistivity $\rho_{yx}$ as a function of magnetic field at selected temperatures. (e),(f) Temperature dependence of the apparent carrier density and mobility, respectively, extracted from the Hall data in (d) using a single-band analysis. The inset in (e) shows a photograph of a typical \ce{CsCr6Sb6} single crystal.}
\label{Fig1}
\end{figure*}

\begin{figure*}
\centering
\includegraphics[scale=0.5]{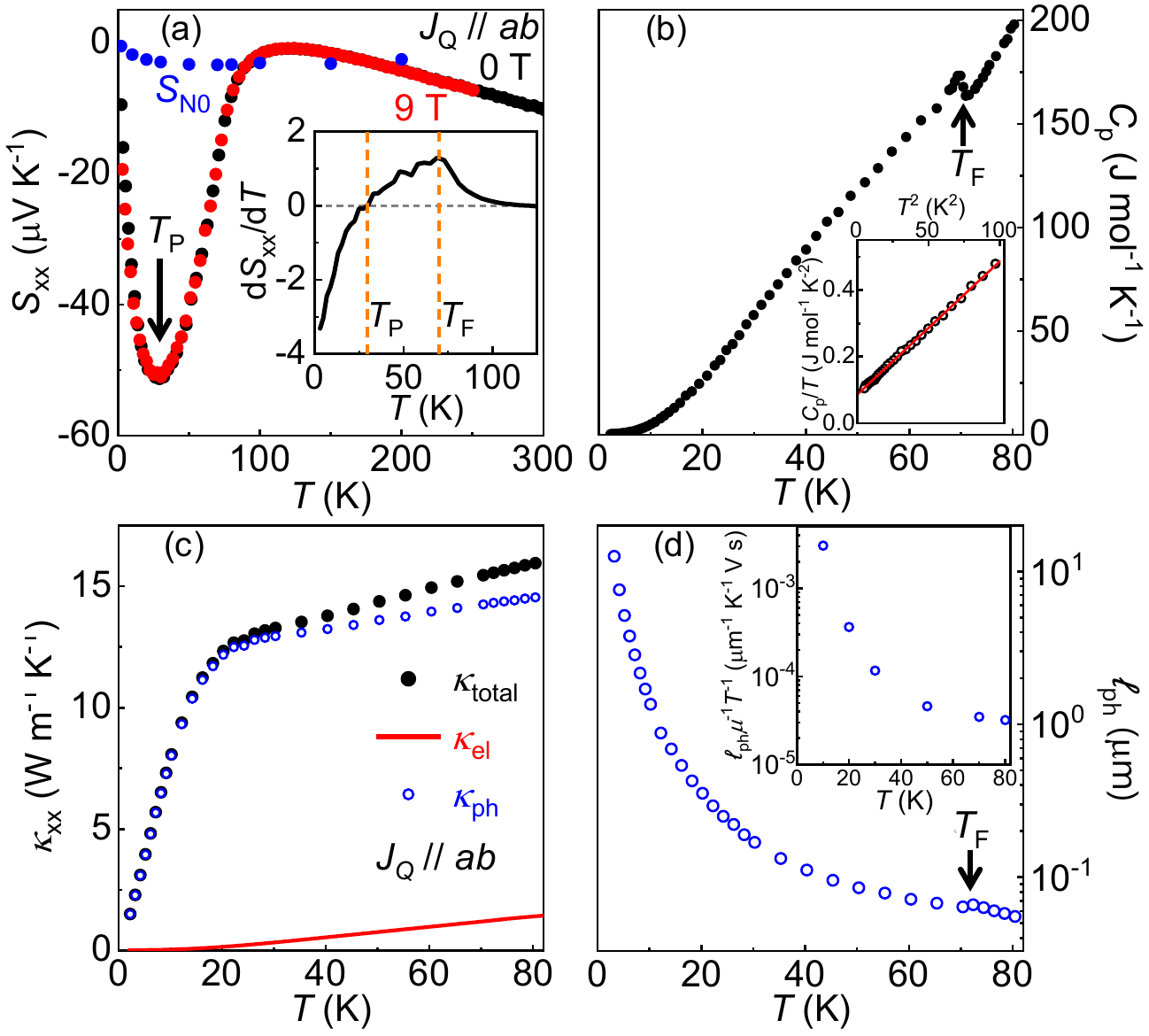}
\caption{(a) Temperature dependence of the Seebeck coefficient $S_{xx}$ of \ce{CsCr6Sb6} measured at 0 and 9 T. The heat current $\mathbf{J_Q}$ was applied within the $ab$ plane, and the magnetic field was directed along the $c$ axis. Blue circles show the calculated conventional diffusion contribution $S_\mathrm{N0}$ without correlation-induced mass renormalization. The inset shows the temperature derivative $dS_{xx}/dT$. (b) Temperature dependence of the specific heat $C_\mathrm{p}$. The inset shows the low-temperature specific heat plotted as $C_\mathrm{p}/T$ versus $T^2$. (c) Temperature dependence of the total thermal conductivity $\kappa_{xx}$ (black solid circles), phonon thermal conductivity $\kappa_\mathrm{ph}$ (blue open circles), and electronic thermal conductivity $\kappa_\mathrm{el}$ (red solid curve). (d) Temperature dependence of the phonon mean free path $\ell_\mathrm{ph}$. The inset shows the temperature evolution of $\ell_\mathrm{ph}\mu^{-1}T^{-1}$.}
\label{Fig2}
\end{figure*}

Single crystals of \ce{CsCr6Sb6} were grown by a self-flux method, as described in Ref.~\cite{CsCrSbNC}. The typical sample dimensions used in this study were approximately $1\times 1\times 0.01~\mathrm{mm}^3$ (see the inset of Fig.~\ref{Fig1}(e)). All measurements were performed in a Physical Property Measurement System (PPMS, Quantum Design Dynacool 9 T). The longitudinal electrical resistivity $\rho_{xx}$ and Hall resistivity $\rho_{yx}$ were measured using a Hall-bar geometry. The Seebeck coefficient $S_{xx}$ and Nernst signal $S_{yx}$ were measured using a one-heater, two-thermometer configuration, as illustrated in Fig.~\ref{Fig2}(a). The longitudinal thermal conductivity $\kappa_{xx}$ was measured using a steady-state method. Below 100 K, errors associated with heat losses were estimated to be less than $\pm 5\%$, while the dominant uncertainty in $\kappa_{xx}$, approximately $\pm 10\%$, arose from the determination of the sample dimensions. Specific heat was measured using the relaxation method. For all transport measurements, the electrical and heat currents were applied within the $ab$ plane, and the magnetic field was applied along the $c$ axis. The transverse signals were obtained by antisymmetrizing the data measured at positive and negative magnetic fields to remove the longitudinal contribution arising from contact misalignment.

\ce{CsCr6Sb6} crystallizes in a van der Waals-like layered trigonal structure with space group $R\overline{3}m$ (No.~166), as shown in Fig.~\ref{Fig1}(a). Its key structural motif is a Cr double-kagome layer extending within the $ab$ plane [Fig.~\ref{Fig1}(b)]. The Sb atoms are closely bonded to the Cr double-kagome network, both within and adjacent to the kagome bilayer, forming a \ce{Cr3Sb3} block. In one unit cell, three \ce{Cr3Sb3} blocks are stacked along the $c$ axis in an $ABC$ sequence, separated by hexagonal Cs layers. The large spacing between adjacent double-kagome layers, approximately 11.559~{\AA} \cite{Zhang2026_CsCr6Sb6}, together with the weak interblock bonding, enables exfoliation down to the monolayer limit \cite{CsCrSbNC}. The electronic states near $E_\mathrm{F}$ are dominated by the Cr double-kagome layers, giving rise to flat bands with an extremely narrow bandwidth of $\sim 40$ meV near the Fermi level \cite{CsCrSbNC,CsCrSbPRB,CsCrSbLiu2025,Zhang2026_CsCr6Sb6,Chen_self}. Moreover, the strong correlations associated with these flat bands have been proposed to stabilize the double-kagome structure itself, leading to a peculiar self-reinforcing effect \cite{Chen_self}. In addition to the flat bands, dispersive bands also appear near $E_\mathrm{F}$, enabling Kondo resonance through the hybridization between localized and itinerant electronic states, as revealed by scanning tunneling spectroscopy (STS) and angle-resolved photoemission spectroscopy (ARPES) measurements \cite{CsCrSbPRB,Zhang2026_CsCr6Sb6}.

The hybridization between flat and dispersive electronic states leaves clear fingerprints in the transport properties of \ce{CsCr6Sb6}. As shown in Fig.~\ref{Fig1}(c), the electrical resistivity exhibits a crossover from high-temperature metallic behavior to a low-temperature insulating-like upturn. A broad kink appears around $T_\mathrm{F}\sim70$ K, close to the temperature below which short-range spin correlations develop in $\mu$SR measurements \cite{CsCrSbLiu2025}. Importantly, near the same temperature, ARPES study revealed an incoherence-to-coherence crossover of the flat-band resonance associated with hybridization \cite{Zhang2026_CsCr6Sb6}.  Upon further cooling below $\sim20$ K, $\rho_{xx}(T)$ displays an approximately logarithmic temperature dependence, indicating that low-energy Kondo scattering becomes prominent \cite{CsCrSbNC}. Below $\sim10$ K, the resistivity deviates from the logarithmic behavior and tends to saturate, consistent with the development of Kondo screening. In contrast, reducing the dimensionality drives a breakdown of the Kondo lattice and induces long-range antiferromagnetic order in thin flakes with thickness below $\sim20$ double-kagome layers \cite{CsCrSbNC}. These dimensionality-tunable phenomena highlight the close competition among flat-band hybridization, Kondo screening, and the Ruderman--Kittel--Kasuya--Yosida (RKKY) interaction.

To estimate the characteristic Kondo scale, we analyze the low-temperature resistivity in more detail. Phenomenologically, the total resistivity ($\rho_{xx}=\rho_\mathrm{n}+\rho_\mathrm{K}$), can be decomposed into a normal metallic background $\rho_\mathrm{n}$ and a Kondo contribution $\rho_\mathrm{K}$. The high-temperature metallic resistivity between 100 and 200 K is nearly linear in temperature [see the upper right inset in Fig.~\ref{Fig1}(c)]. The normal contribution is thus estimated by
$\rho_\mathrm{n}=\rho_0+AT$, where $\rho_0$ is the residual resistivity.  The Kondo contribution is modeled using the Hamann expression \cite{Hamann}, which provides an approximation to Nagaoka's self-consistent treatment \cite{Nagaoka}. Although the original Hamann approximation underestimates the impurity-spin contribution for $T<T_\mathrm{K}$, this limitation can be partly mitigated by introducing an effective temperature $T_\mathrm{eff}=\sqrt{T^2+T_\mathrm{W}^2}$, where $k_\mathrm{B}T_\mathrm{W}$ represents an effective RKKY interaction scale \cite{Liang_ef,Kastner_ef,Ding_ef}. Within this modified Hamann framework, the Kondo resistivity is written as \cite{Hamann,Liang_ef,Kastner_ef,Ding_ef}:
\begin{equation}
\rho_\mathrm{K}(T_\mathrm{eff})=\rho_1
\left[
1-\frac{\ln\left(T_\mathrm{eff}/T_\mathrm{K}\right)}
{\sqrt{\ln^2\left(T_\mathrm{eff}/T_\mathrm{K}\right)+S(S+1)\pi^2}}
\right],
\label{eq:rhoK}
\end{equation}
where $\rho_1$ is a constant and $S$ is an effective spin parameter. The Hamann expression was originally derived for spin-$1/2$ impurities. In practical analyses of experimental data, however, $S$ is often treated as a phenomenological fitting parameter, typically taking values smaller than $1/2$ \cite{Sutsch_Sfree,Zeng2021_Sfree,Ding_ef}. As shown by the red curve in Fig.~\ref{Fig1}(c), the resistivity below 70 K can be well described by $\rho_{xx}=\rho_\mathrm{0}+AT+\rho_\mathrm{K}$, yielding the fitting parameters $\rho_0=0.27$ $\mathrm{m}\Omega$ cm, $A=1.2 \times 10^{-3}$ $\mathrm{m}\Omega$ cm K$^{-1}$, $\rho_1=0.94$ $\mathrm{m}\Omega$ cm, $T_\mathrm{K}=15$ K, $T_\mathrm{W}=2.3$ K, and $S=0.1$. Notably, below $\sim 4$ K ($\ll T_\mathrm{K}$), the modified Hamann formula also captures the $-T^2$ dependence of the resistivity [lower left inset of Fig.~\ref{Fig1}(c)], a characteristic low-temperature behavior expected from Nagaoka's theory of the single-impurity Kondo effect \cite{Nagaoka,Kastner_ef,Ding_ef,Nozieres1974_T2,Gauzzi_T2}. We emphasize that the fitted $T_\mathrm{K}=15$ K represents a low-energy Kondo scattering scale within this phenomenological description. It should not be identified with the onset of electronic coherence, which occurs at the distinctly higher temperature $T_\mathrm{F}\sim70$ K according to the ARPES measurements \cite{Zhang2026_CsCr6Sb6}.

Figure~\ref{Fig1}(d) shows the Hall resistivity $\rho_{yx}$ of \ce{CsCr6Sb6}. Over the measured temperature range, $\rho_{yx}$ varies nearly linearly with magnetic field and exhibits a negative slope, indicating that the charge transport is dominated by electron-like carriers. The apparent carrier density and mobility extracted from a single-band analysis are plotted in Figs.~\ref{Fig1}(e) and \ref{Fig1}(f), respectively. Upon cooling, the carrier density decreases continuously, while the mobility increases smoothly. Compared with the structurally related kagome metal \ce{CsV6Sb6} \cite{Yin_CsV6Sb6,Shi_CsV6Sb6}, \ce{CsCr6Sb6} possesses a much lower carrier density of order $\sim 10^{20}$ cm$^{-3}$ and a relatively higher mobility reaching $\sim 50$ cm$^2$ V$^{-1}$ s$^{-1}$ at low temperature. These features are consistent with a low carrier density electronic state emerging from flat-band-derived Kondo hybridization.

Figure~\ref{Fig2}(a) presents the Seebeck coefficient $S_{xx}$ of \ce{CsCr6Sb6}. Over the entire measured temperature range below 300 K, $S_{xx}$ remains negative, consistent with electronlike transport inferred from the Hall measurements. Upon cooling from room temperature, the magnitude of $S_{xx}$ first decreases gradually down to about 100 K and then begins to increase. The most pronounced change occurs around $T_\mathrm{F}$, as evidenced by a peak in $dS_{xx}/dT$ [inset of Fig.~\ref{Fig2}(a)]. This temperature coincides with both the development of short-range magnetic correlations and the incoherence-to-coherence crossover of the flat-band resonance  \cite{CsCrSbLiu2025,Zhang2026_CsCr6Sb6}. The onset of the enhanced $S_{xx}$ is therefore most likely associated with the hybridization-driven reconstruction of the low-energy quasiparticle state. With further cooling, a pronounced negative minimum develops in $S_{xx}$ around $T_\mathrm{p}\sim25$ K.

In general, the measured thermopower $S_{xx}=S_\mathrm{N}+S_\tau+S_\mathrm{ph}$ may contain a conventional band-diffusion
term $S_\mathrm{N}$, an energy-dependent scattering contribution $S_\tau$, and
a phonon-drag contribution $S_\mathrm{ph}$,\cite{Neto_MoS2}. To estimate $S_\mathrm{N}$, we solve the linearized Boltzmann transport equation within an effective
three-dimensional (3D) single parabolic band  \cite{Neto_MoS2,Pichanusakorn_BTE}
\begin{equation}
S_\mathrm{N}
=
-\frac{k_\mathrm{B}}{e}
\left[
\frac{
\left(r+\frac{5}{2}\right)F_{r+\frac{3}{2}}(\eta)}
{
\left(r+\frac{3}{2}\right)F_{r+\frac{1}{2}}(\eta)}
-\eta
\right],
\label{eq:SN_BTE}
\end{equation}

where $e>0$ is the elementary charge, $k_\mathrm{B}$ is the Boltzmann constant, $\eta$ is the reduced Fermi potential, and
$F_j(\eta)=\int_0^\infty x^j/[1+\exp(x-\eta)]\,dx$ is the Fermi integral.
We take the scattering parameter $r=0$ in the constant relaxation time approximation $\tau(E)=\tau_0E^r$. At each temperature, $\eta$ is determined from the measured Hall carrier density
using

\begin{equation}
n(T)=
\frac{1}{2\pi^2}
\left(
\frac{2m_\mathrm{b}k_\mathrm{B}T}{\hbar^2}
\right)^{3/2}
F_{1/2}(\eta),
\end{equation}

We first take $m_\mathrm{b}=m_0$, where $m_0$ is the free-electron mass, to obtain an unrenormalized single-band reference without correlation-induced mass renormalization. As shown in Fig.~\ref{Fig2}(a), the resulting $S_\mathrm{N0}$ is of the same order of magnitude as the measured thermopower in the high-temperature normal state regime. Below 80 K, however, $S_\mathrm{N0}$ remains weakly temperature dependent and is only a few \textmu V K$^{-1}$ around 20--30 K, whereas the measured $|S_{xx}|$ is strongly enhanced. This indicates that the bare-band reference alone cannot account for the low-temperature thermopower. To examine the influence of quasiparticle mass renormalization, we then repeat the calculation using $m_\mathrm{b}=m^*=60m_0$, as inferred from the low-temperature Sommerfeld coefficient (see discussion below). Assuming a temperature-independent $m^*$ yields a strongly enhanced diffusion thermopower, with $S^*_\mathrm{N}\approx-170$ \textmu V K$^{-1}$ at 20 K, substantially exceeding the measured $S_{xx}$. This comparison demonstrates that the diffusion thermopower is highly sensitive to quasiparticle mass renormalization. However, directly applying the low-temperature thermodynamic effective mass over the entire temperature range is also an oversimplification. In addition, the single parabolic band and constant relaxation time approximations introduce further uncertainties, particularly in the hybridized electronic structure of \ce{CsCr6Sb6}. We therefore regard $S_\mathrm{N0}$ and $S^*_\mathrm{N}$ as two limiting estimates illustrating the sensitivity of $S_\mathrm{N}$ to mass renormalization, rather than as quantitative determinations of $S_\mathrm{N}$.

A sizable phonon-drag contribution may become important at lower temperatures. Phonon drag is commonly most pronounced at temperatures of order $\Theta_\mathrm{D}/10$--$\Theta_\mathrm{D}/5$. As shown in the inset of Fig.~\ref{Fig2}(b), fitting the low-temperature specific heat to $C_\mathrm{p}=\gamma T+\beta_\mathrm{C}T^3$ gives $\gamma=90$ mJ mol$^{-1}$ K$^{-2}$ and a Debye temperature $\Theta_\mathrm{D}=185$ K, obtained from $\beta_\mathrm{C}=12\pi^4NR/(5\Theta_\mathrm{D}^3)$, where $N$ is the number of atoms per formula unit and $R$ is the ideal-gas constant. The corresponding average sound velocity is $v_s=1781$ m s$^{-1}$, estimated from $v_s=\Theta_\mathrm{D}(k_\mathrm{B}/\hbar)(6\pi^2N/V)^{-1/3}$, where $V$ is the volume per formula unit. The characteristic phonon-drag window is therefore approximately 18--37 K, directly overlapping the broad Seebeck minimum near 25 K. In addition,  the characteristic thermal-phonon wave vector can be estimated as $q\sim k_\mathrm{B}T/(\hbar v_s)$, whereas the Fermi wave vector is $k_\mathrm{F}=(3\pi^2n)^{1/3}$. Using the measured temperature-dependent carrier density, we obtain $q/(2k_\mathrm{F})\approx0.33$ and 0.43 at 20 and 30 K, respectively, indicating appreciable momentum-transfer phase space for electron--phonon scattering. By contrast, $q/(2k_\mathrm{F})$ decreases to $\sim0.20$ at 10 K and $\sim0.05$ at 2 K. Together with the rapidly vanishing thermal-phonon population, this strongly suppresses the phonon-drag contribution as $T\rightarrow0$.

A commonly used phenomenological estimation of $S_\mathrm{ph}$ is the Herring relation,
$S_\mathrm{ph}=\beta_\mathrm{ep}v_s\ell_s/(\mu T)$, where
$0\leq\beta_\mathrm{ep}\leq1$ characterizes the efficiency of phonon momentum transfer to charge carriers
\cite{Neto_MoS2,Herring,FeSb2PhononDrag}.
The phonon mean free path $\ell_s$ relevant to phonon drag is commonly approximated by the thermal mean free path $\ell_\mathrm{ph}$, which can be estimated for a 3D system from
$\ell_\mathrm{ph}=3\kappa_\mathrm{ph}/(C_\mathrm{ph}v_s)$. Here, $\kappa_\mathrm{ph}$ is the phonon thermal conductivity and $C_\mathrm{ph}=C_\mathrm{p}-\gamma T$ is the phonon specific heat.
As shown in Fig.~\ref{Fig2}(c), the measured thermal conductivity is dominated by $\kappa_\mathrm{ph}$. The electronic part is estimated from the Wiedemann--Franz law,
$\kappa_\mathrm{el}=L_0T\sigma_{xx}$, where $L_0$ is the Lorenz number and $\sigma_{xx}$ is the longitudinal electrical conductivity. Any additional magnetic heat-current contribution associated with short-range correlations would introduce further uncertainty into the extracted $\ell_\mathrm{ph}$. Interestingly, $\kappa_\mathrm{ph}$ lacks the conventional crystalline peak associated with the crossover between boundary and phonon--phonon Umklapp scattering. Instead, it displays an  glasslike behavior with nearly linear temperature dependence above 20 K, suggesting strong phonon scattering that may be coupled to the short-range magnetic correlations
\cite{Li_glass,Sharam_glass,Uehara2022_glass,Casto_glass,Yang_glass,yang2023_glass}.
The resulting $\ell_\mathrm{ph}$ and $\ell_\mathrm{ph}/(\mu T)$ evolve smoothly through 20--25 K [inset of Fig.~\ref{Fig2}(d)]. However, $\ell_\mathrm{ph}$ represents an average over heat-carrying phonons and need not be identical to $\ell_s$. Moreover, $\beta_\mathrm{ep}$ may be temperature dependent and is not directly determined by the present measurements. Therefore, $\ell_\mathrm{ph}/(\mu T)$ cannot be identified quantitatively with $S_\mathrm{ph}$. We conclude that phonon drag is kinematically plausible and may contribute substantially near the Seebeck minimum, but its magnitude cannot be uniquely determined from the available data.

The weak magneto-Seebeck response is also compatible with a sizable phonon-drag contribution. As shown in Fig.~\ref{Fig2}(a), the $S_{xx}(T)$ curves measured at 0 and 9 T nearly overlap, except below $\sim10$ K. Nevertheless, the absence of a sizable response at 9 T does not by itself exclude $S_\tau$ contributions from correlated quasiparticles. Spectroscopic measurements of \ce{CsCr6Sb6} reported a hybridization gap of approximately 10--15 meV \cite{CsCrSbNC}. For comparison, the Zeeman energy at 9 T is only $g\mu_\mathrm{B}B\approx1$ meV for $g=2$, more than an order of magnitude smaller than the hybridization energy scale. Thus, 9 T may be insufficient to substantially perturb the correlated electronic state associated with hybridization. Weak magnetothermopower at comparable laboratory fields has also been reported in other Kondo systems. For example, in the Kondo insulator \ce{CeRu4Sn6}, whose characteristic gap is much smaller than that of \ce{CsCr6Sb6}, a 9 T field has little effect on the Seebeck response \cite{Bruning_CeRuSn}. Similarly, no obvious field dependence of the Seebeck signal  is reported in \ce{CeCu2Si2} up to 7 T \cite{Sun_CeRuSi}. Therefore, the rapid enhancement in $S_{xx}$ below $T_\mathrm{F}$ is very likely associated with the formation of the correlated quasiparticle state, whereas the low-temperature extrema near $T_\mathrm{p}$ may contain mixed contributions from $S_\mathrm{ph}$ and  $S_\tau$.

\begin{figure*}
\centering
\includegraphics[scale=0.7]{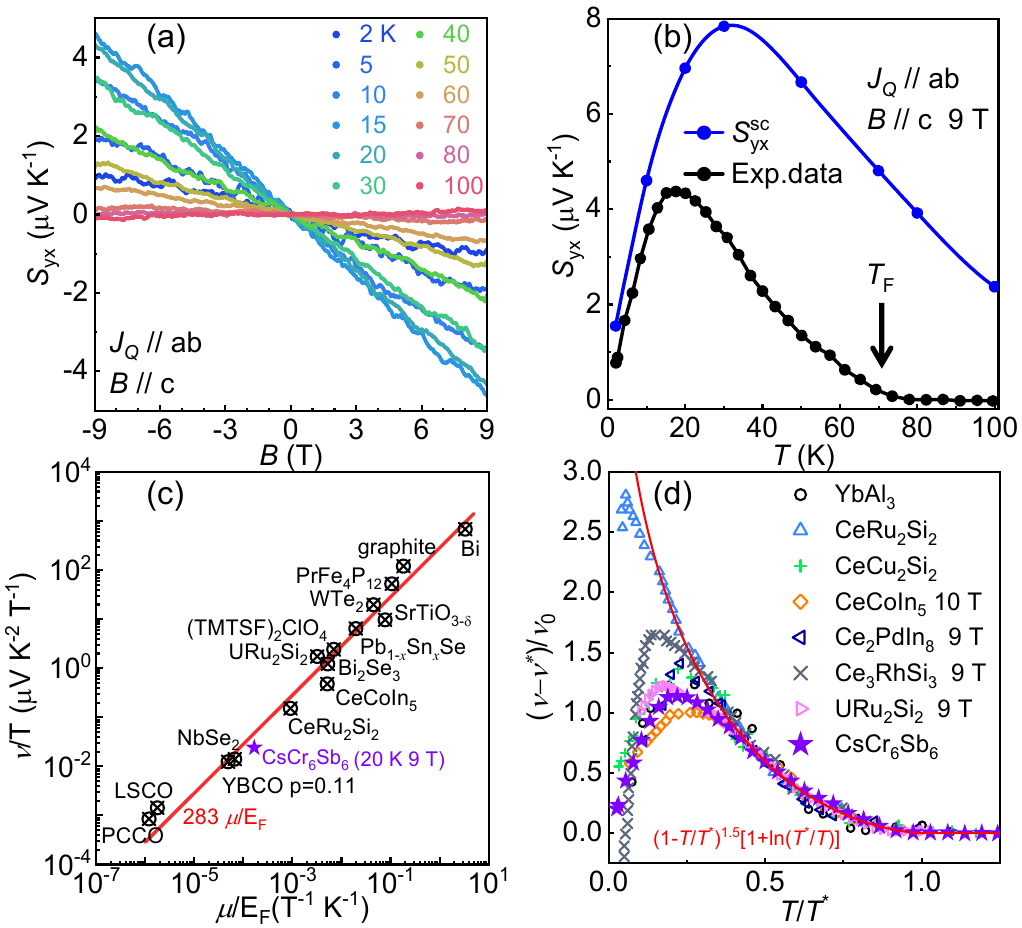}
\caption{(a) Magnetic-field dependence of the Nernst signal $S_{yx}$ at selected temperatures. (b) Temperature dependence of $S_{yx}$ measured at 9 T. The semiclassical $S^\mathrm{sc}_{yx}$ estimated from Eq.~\ref{eq:nernst} using the temperature-dependent mobility and Fermi energy is shown for comparison. (c) Semiclassical scaling of $\nu/T$ with $\mu/E_\mathrm{F}$ for various materials \cite{behnia2009,behnia2016}, where $\nu=S_{yx}/B$. The red solid line represents Eq.~\ref{eq:nernst}, while the purple star denotes \ce{CsCr6Sb6} at 20 K, using $E_\mathrm{F}/k_\mathrm{B}\sim36$ K. (d) Universal scaling of the Nernst coefficient according to Eq.~\ref{eq:universal} (red solid line), compared with various prototypical heavy-fermion compounds \cite{Yang_universial,behnia2009,Sun_CeRuSi,Wei_YbAl3,CeRhSi,CePdIn8,CeCoIn5,URu2Si2}.}
\label{Fig3}
\end{figure*}

The transverse thermoelectric response provides a complementary probe of the electronic reconstruction associated with flat band resonance and hybridization below $T_\mathrm{F}$. As shown in Fig.~\ref{Fig3}(a), the Nernst signal $S_{yx}=E_y/\nabla_xT$ varies nearly linearly with magnetic field up to 9 T at all measured temperatures. Above $T_\mathrm{F}$, $S_{yx}$ is nearly negligible. Upon cooling below this temperature, it increases rapidly and develops a maximum around 20 K [Fig.~\ref{Fig3}(b)]. The rapid enhancement of $S_{yx}$ below $T_\mathrm{F}$ suggests that the Nernst signal is strongly modified by the emergence of correlated quasiparticles. Although a phonon-drag contribution may generate a peak-like Nernst response, it is unlikely to be the dominant contribution, particularly over the intermediate-temperature range of 30--80 K. As shown in Fig.~\ref{Fig3}(c), the Nernst coefficient $\nu=S_{yx}/B$ falls close to the semiclassical scaling relation \cite{behnia2009,behnia2016}:
\begin{equation}
\frac{\nu}{T}=
\frac{\pi^2 k_\mathrm{B}^2}{3e}
\frac{\mu}{E_\mathrm{F}}
\simeq
283\frac{\mu}{E_\mathrm{F}}.
\label{eq:nernst}
\end{equation}
 The Fermi energy $E_\mathrm{F}$ is estimated using $E_\mathrm{F}=\hbar^2(3\pi^2 n)^{2/3}/2m^{*}$, where $n$ is the carrier density obtained from the Hall data. The effective electron mass $m^{*}$ can be estimated from $m^{*}/m_0=\gamma/\gamma_0$.  Within a 3D single parabolic band approximation, the free-electron Sommerfeld coefficient is estimated as $\gamma_0=\frac{1}{3}\pi^2k_\mathrm{B}^2N(E_\mathrm{F})VN_\mathrm{A}$. Here, $N(E_\mathrm{F})=m_0k_\mathrm{F}/(\pi^2\hbar^2)$ is the free-electron DOS at $E_\mathrm{F}$, $V=306.6$ $\mathrm{\AA}^3$/f.u. is the volume per formula unit, and $N_\mathrm{A}$ is Avogadro's constant. Using the carrier density $n=1.3\times10^{20}$ cm$^{-3}$ measured at 2 K, we obtain $\gamma_0\approx1.5$ mJ mol$^{-1}$ K$^{-2}$ and $m^{*}\approx60m_0$, in good agreement with the previous report \cite{CsCrSbNC}. Using this effective mass together with the temperature-dependent carrier density, we estimate $E_\mathrm{F}\sim18$ K and $\sim36$ K at $T=2$ and 20 K, respectively. 

The semiclassical scaling in Eq.~\ref{eq:nernst} also applies to strongly correlated systems, including cuprates and heavy-fermion materials [Fig.~\ref{Fig3}(c)] \cite{behnia2009,behnia2016}. In heavy-fermion systems, the enhanced effective mass substantially reduces the effective Fermi energy, with $E_\mathrm{F}\sim k_\mathrm{B}T_\mathrm{K}$ \cite{Behnia_2004}. Correlation effects can therefore enter the semiclassical expression through the renormalized quasiparticle parameters. Using the temperature-dependent $\mu(T)$ and $E_\mathrm{F}(T)$, we estimate the semiclassical Nernst signal $S^\mathrm{sc}_{yx}$ assuming a temperature-independent $m^*$. As shown in Fig.~\ref{Fig3}(b), $S^\mathrm{sc}_{yx}$ reproduces the order of magnitude and the overall broad low-temperature enhancement of the measured Nernst signal. 
Note that the semiclassical relation in Eq.~\ref{eq:nernst} is most appropriately applied to the low-temperature regime \cite{behnia2009,behnia2016}. Indeed, above $\sim20$ K, the simple semiclassical estimation overestimates the measured Nernst signal and fails to reproduce its detailed temperature dependence. In particular, above about 80 K, the measured signal becomes nearly negligible, whereas the estimated value remains sizable. Similar to the estimation of $S^*_\mathrm{N}$, large uncertainties may arise from the simplified evaluation of $E_\mathrm{F}$, the assumption of a temperature-independent effective mass, and the neglect of additional energy-dependent scattering. Nevertheless, the order of magnitude agreement between the estimated $S^\mathrm{sc}_{yx}$ and the measured $S_{yx}$ supports a substantial electronic quasiparticle contribution to the Nernst enhancement, associated with the small renormalized $E_\mathrm{F}$ and enhanced mobility of the correlated state.

Since phonon drag may be appreciable within approximately 18--37 K, we now focus on the intermediate-temperature range of 30--80 K. While the semiclassical relation provides an order of magnitude estimation of the Nernst signal, the universal scaling identified by Yang \textit{et al.} provides a complementary description of its temperature dependence over a broad intermediate-temperature range in various heavy-fermion systems \cite{Yang_universial}:
\begin{equation}
    \frac{\nu-\nu^{*}}{\nu_0}
    =\left(1-\frac{T}{T^{*}}\right)^{3/2}
    \left[1+\ln\left(\frac{T^{*}}{T}\right)\right],
    \label{eq:universal}
\end{equation}
where $\nu^{*}$ is the nearly temperature-independent high-temperature background, $\nu_0$ is a temperature-independent normalization constant, and $T^{*}$ is the onset temperature below which $\nu$ departs from $\nu^{*}$.  As shown in Fig.~\ref{Fig3}(d), the Nernst coefficient of \ce{CsCr6Sb6} between 30 and 80 K can also be well reproduced by the universal scaling curve with $T^{*}=80$ K, $\nu^{*}=-1.22\times10^{-3}$ \textmu V K$^{-1}$ T$^{-1}$ and $\nu_0=0.43$ \textmu V K$^{-1}$ T$^{-1}$. Importantly, $T^{*}\approx80$ K is close to the incoherence-to-coherence crossover of the flat-band resonance directly observed by ARPES experiments \cite{Zhang2026_CsCr6Sb6}.  This suggests that the enhanced Nernst signal in the intermediate-temperature range of 30--80 K is closely tied to the emergence of the correlated quasiparticle state. Although the microscopic origin of the universal scaling remains unclear, it is likely related to an asymmetric heavy-quasiparticle DOS associated with hybridization \cite{Yang_universial}. Indeed, an asymmetric Fano line shape in \ce{CsCr6Sb6} has been revealed by STS measurements \cite{CsCrSbPRB}. Within the two-fluid framework, the magnetic susceptibility $\chi_h(T)$ of the emergent heavy-electron component is predicted to exhibit the same universal temperature dependence as Eq.~\ref{eq:universal} \cite{Yang_universial,Yang_twofluid}. In the phenomenological picture proposed in Ref.~\cite{Yang_universial}, combining skew scattering with an energy-over-temperature scaling of the heavy-quasiparticle spectrum leads to $\nu(T)\propto\chi_h(T)$, providing a possible explanation for the observed Nernst scaling. The predicted temperature dependence of $\chi_h(T)$ has received experimental support in a number of heavy-fermion compounds, particularly from the Knight-shift anomaly \cite{Yang_knight,Curro_knight,Nakatsuji_knight}. It would therefore be interesting to further track an analogous Knight-shift anomaly in \ce{CsCr6Sb6} through NMR measurements. 
Note that although both the semiclassical relation in Eq.~\ref{eq:nernst} and the universal temperature scaling in Eq.~\ref{eq:universal} provide useful phenomenological descriptions of the Nernst response in \ce{CsCr6Sb6} and in other heavy-fermion systems, including \ce{CeRu2Si2}, \ce{CeCoIn5}, and \ce{URu2Si2} [Figs.~\ref{Fig3}(c,d)], the underlying microscopic mechanisms remain unresolved and warrant further investigation.

Finally, we note that a lambda-like anomaly appears in the specific heat at $T_\mathrm{F}$, which might conventionally be associated with the onset of long-range magnetic order. However, muon spin relaxation measurements reveal only short-range magnetic correlations below $T_\mathrm{F}$ \cite{CsCrSbLiu2025}. Similar specific-heat anomalies without conventional long-range magnetic order have been reported in frustrated magnets, such as $\gamma$-\ce{BaCo2(PO4)2} and \ce{Yb2Ti2O7} \cite{Nair_short,Hodges_short}. The anomaly in \ce{CsCr6Sb6} may therefore be associated with short-range magnetic ordering in the frustrated kagome lattice, while its microscopic origin remains to be clarified.

In summary, we have investigated the electrical, thermoelectric, and thermal transport properties of the flat-band-derived kagome Kondo system \ce{CsCr6Sb6}, in which Kondo physics originates from hybridization between flat and dispersive bands. Both the Seebeck coefficient and Nernst signal begin to increase rapidly below $T_\mathrm{F}\sim70$ K, close to the incoherence-to-coherence crossover of the  flat-band resonance observed by previous ARPES experiments. The Nernst coefficient in the intermediate-temperature range of approximately 30--80 K follows the universal scaling reported for various heavy-fermion materials, while its low-temperature magnitude is consistent with strongly renormalized quasiparticles characterized by a small effective Fermi energy and enhanced mobility. At lower temperatures, the Seebeck and Nernst responses develop broad extrema around 20--25 K, where phonon drag and correlated quasiparticles may both contribute.

\section*{Acknowledgments}

We thank Guiwen Wang and Yan Liu at the Analytical and Testing Center of Chongqing University for technical support. This work has been supported by National Natural Science Foundation of China (Grants Nos. 12474141, 52522201), Natural Science Foundation of Chongqing, China (Grant No. CSTB2025NSCQ-GPX0729), Fundamental Research Funds for the Central Universities, China (2025CDJ-IAISYB-034), the Venture and Innovation Support Program for Chongqing Overseas Returnees (Grant No. cx2024007), and Chinesisch-Deutsche Mobilitätsprogamm of Chinesisch-Deutsche Zentrum für Wissenschaftsförderung (Grant No. M-0496). Y.C. acknowledges the support by the National Natural Science Foundation of China (Grant Nos. 12374081, 12227806), the Open Research Fund of the Pulsed High Magnetic Field Facility (Grant No. WHMFC2024007), and Huazhong University of Science and Technology.

\end{document}